# Artificial neural networks for predicting the viscosity of lead-containing glasses

Patrick dos Anjos[1]*
Lucas de Almeira Quaresma*
Marcelo Lucas Pereira Machado*

*Federal Institute of Espírito Santo (IFES), Vitória, ES, Brazil

## Abstract

The viscosity of lead-containing glasses is of fundamental importance for the manufacturing process, and can be predicted by algorithms such as artificial neural networks. The SciGlass database was used to provide training, validation and test data of chemical composition, temperature and viscosity for the construction of artificial neural networks with node variation in the hidden layer. The best model built with training data and validation data was compared with 7 other models from the literature, demonstrating better statistical evaluations of mean absolute error and coefficient of determination to the test data, with subsequent sensitivity analysis in agreement with the literature. Skewness and kurtosis were calculated and there is a good correlation between the values predicted by the best neural network built with the test data.

## 1 Introduction

Lead-containing glasses are materials that have a high refractive index, with greater ease of cutting and polishing when compared to other types of glass. They are applied in crystal glasses, also applied in the development of artistic glasses and in industrial environments due to their flexibility [1].

The viscosity of glasses is an important parameter in their manufacturing process. Viscosity variation governs the temperature of processes such as working point, littleton softening point, annealing point, glass transition and strain point [1], as indicated in Table 1. Viscosity can be measured experimentally by methods such as rotational viscometer [2] or predicted by algorithms through a database [3], such as the artificial neural network.

The aim of this work is to model the viscosity as a function of the chemical composition and the temperature of lead-containing glasses through artificial neural networks intermediated by a database.

---

1  E-mail: patrick.dosanjos@outlook.com

Table 1. Reference temperatures and related viscosity.

| Reference temperature | Viscosity (Pa.s) |
|---|---|
| Working point | $1-10^3$ |
| Littleton Softening point | $10^{6.6}$ |
| Annealing point | $\sim 10^{12}$ |
| Glass Transition | $\sim 10^{12}-10^{12.5}$ |
| Strain point | $\sim 10^{13.5}$ |

## 2  Materials and Methods

The database used was SciGlass (https://github.com/epam/SciGlass), previously used to predict other properties of chemical systems [4]. The preprocessing of the database was established through the conditions:

I) Missing values (NaN) converted to 0;
II) $SiO_2+Al_2O_3+B_2O_3+CaO+K_2O+Na_2O+PbO = 100$ and $SiO_2 > 0$;
III) Viscosity $< 10^{12}$ Pa.s.

The preprocessed database provided data on chemical composition (%mol) and temperature (K), the predictor variables, and viscosity ($\eta$, in Pa.s), the variable to be predicted. In Figure 1 it is possible to observe the dispersion of viscosity data in relation to temperature in the preprocessed database, since the viscosity has a variation of approximately 13 orders of magnitude ($\approx 0.1$-$<10^{12}$ Pa.s). Viscosity can be modeled mathematically through linear regression [2] and by artificial neural networks [3]. Artificial neural networks have advantages because they provide greater capacity for learning in complex relationships between variables, also in noisy data [3], in addition to presenting the universal approximation theorem.

The universal approximation theorem states that artificial neural networks with 1 hidden layer approximate continuous real functions defined in a compact subset with arbitrary precision [5]. Thus, the theorem describes that an artificial neural network with 1 hidden layer can be built with $d \leq k^{3/2}$ nodes where $k = n + 4$ and $n$ is equivalent to input variables dimension [6].

Data were normalized between 0 and 1 and separated into training, validation and test data. Test data were **not** used during the training of artificial neural networks. Forty-one different artificial neural networks were constructed with statistical evaluations of mean absolute error (MAE) and coefficient of determination ($R^2$) in relation to the test data, with the EarlyStopping technique to avoid over/underfitting of the model.

The MAE, $R^2$ results in relation to the test data were compared with the S2, Watt-Fereday, Bomkamp, Lakatos and Riboud models described in Vargas et al. (2001) and Duchesne and ANNliq models, described by Duchesne et al. (2013). Sensitivity analysis was performed using the connection weights method [7] and analysis of kurtosis and asymmetry of deviations in relation to the test data.

## 3  Results and discussion

The best artificial neural network, with lowest MAE and highest $R^2$, has 26 nodes in the hidden layer. The loss curve (Mean Squared Error Loss) in relation to the epochs (maximum 10K), together with the validation score curve ($R^2$) of the best artificial neural network can be seen in Fig. 2.

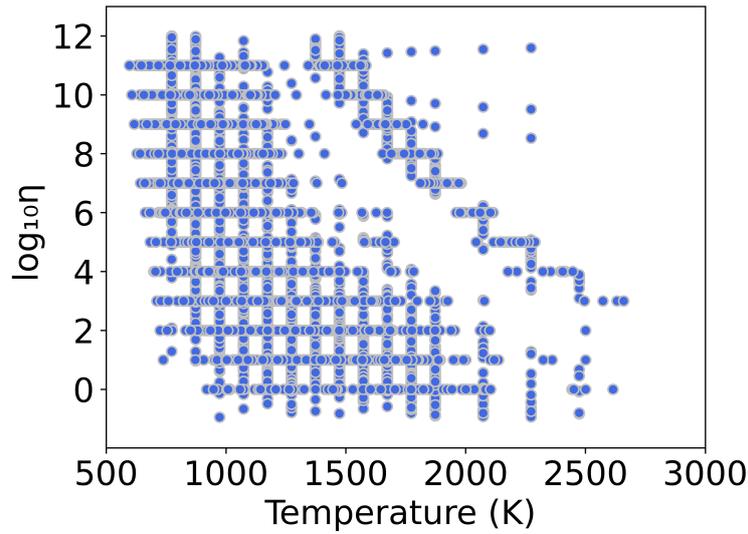

Figure 1. Viscosity (log₁₀ η) versus Temperature preprocessing data.

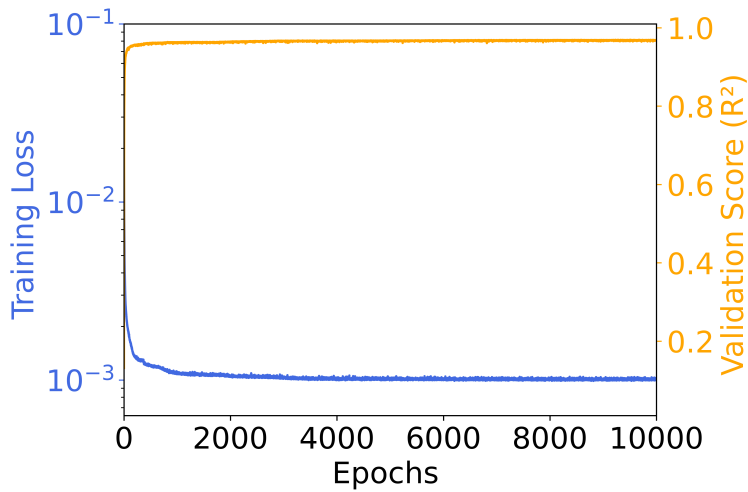

Figure 2. Training Loss (Mean Squared Error) and Validation Score (Coefficient of determination — $R^2$) in the ANN-26 hidden nodes.

The mean absolute error (MAE) and standard deviation of deviation (STD) results in relation to the test data in the S2, Watt-Fereday, Bomkamp, Lakatos, Riboud, Duchesne, ANNliq models and in the present work are represented in Table 2.

The model of the present work has lower MAE and STD in relation to the 7 other models in the literature, indicating better effectiveness in relation to the test data. The sensitivity analysis (Fig. 3) of the model of the present work indicates the values of relative importance in numerical values in the bar graph and in percentage (%) in the inscribed table.

The relative importance in numerical values indicates that their absolute value of the input variable concerns greater or lesser variations in the output variable when the input variable varies [8]. For example, the input variable T (Temperature) has greater sensitivity in relation to all other input variables, as it has a higher absolute value (7.09).

Table 2. MAE (log₁₀ η) and STD (log₁₀ η) values in the studied viscosity models (Mod.)

| Mod. | S2 | Watt-Fereday | Bomkamp | Lakatos | Riboud | Duchesne | ANNLiq | Present work |
|------|-------|--------------|---------|---------|--------|----------|--------|--------------|
| MAE  | 3.402 | 19.257       | 21.362  | 3.512   | 2.794  | 3.252    | 3.358  | 0.352        |
| STD  | 2.016 | 24.326       | 25.704  | 4.012   | 2.238  | 2.353    | 2.387  | 0.415        |

The sign of relative importance indicates that, if the sign of relative importance is positive, the input variable is directly proportional to the output variable, and when the sign of relative importance is negative, there is an inversely proportional relationship to the output variable [8]. The addition of CaO in a chemical system lowers its viscosity at a given temperature (-1.96).

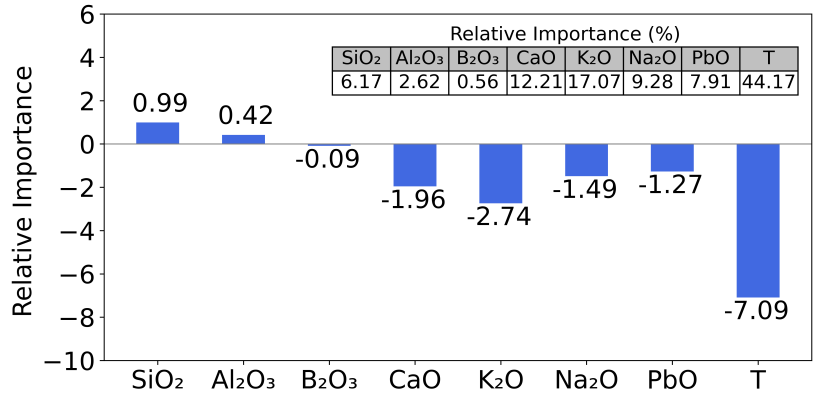

Figure 3. Relative Importance of the input variables to output variable in ANN-26 hidden nodes.

The relative importance in percentage values shows the weight of each input variable in relation to the output variable. The variation of $B_2O_3$ indicates that it does not significantly affect the output variable, with the input variable having the lowest percentage value of relative importance (0.56%). The addition of silica ($SiO_2$) to silicates causes an increase in viscosity, as the silicon forms network-forming ions. Alumina ($Al_2O_3$) can also be considered a network-forming species, increasing the viscosity of silicates, requiring an $M^+$ or $M^{2+}$ atom for its electronic stabilization [2].

Chemical species composed of alkaline and/or alkaline earth elements are network-modifying species, breaking and weakening silicate networks, respectively, and therefore decreasing the viscosity of a chemical system at a given temperature. The Equation 1 and Equation 2 indicate the mechanisms of viscosity reduction by adding oxides composed of alkaline elements and alkaline earth elements, respectively [2].

The chemical species CaO, $K_2O$, $Na_2O$ have alkaline or alkaline earth chemical elements and PbO also decreases the viscosity of a chemical system at a given temperature [9].

$$\equiv Si\text{-}O\text{-}Si\equiv\ +\ M_2O \rightarrow\ \equiv Si\text{-}O^-M^+ +\ M^+O^-Si\equiv \tag{1}$$

$$\equiv Si\text{-}O\text{-}Si\equiv\ +\ MO \rightarrow\ \equiv Si\text{-}O\text{-}M\text{-}O\text{-}Si\equiv \tag{2}$$

$B_2O_3$ is classified as an amphoteric oxide, which can decrease or increase the viscosity of chemical systems and temperature is inversely proportional to viscosity as described by the Vogel-Fulcher-Tammann (VFT) equation (Equation 3), where a, b and c are constant, T is the temperature (K) and $\eta$ is the viscosity [2].

$$\log \eta = a + b/(T-c) \tag{3}$$

The relative importance of the input variables $SiO_2$, $Al_2O_3$, $B_2O_3$, CaO, $K_2O$, $Na_2O$, PbO and temperature (T) performed by the connection weights method [7] are in agreement with the literature. The histogram of the deviations of the values predicted by the model of the present work in relation to the test data (Figure 4) indicate a negative asymmetry, with a positive kurtosis.

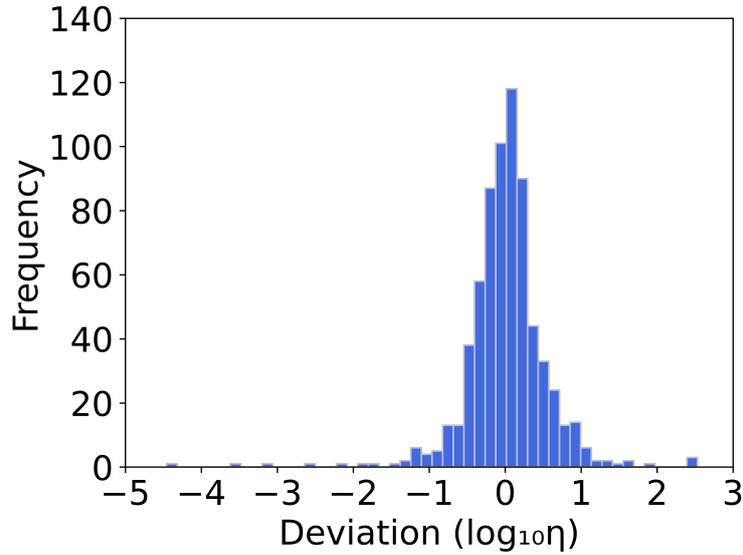

Figure 4. Histogram of deviation between predicted values of present work to test data.

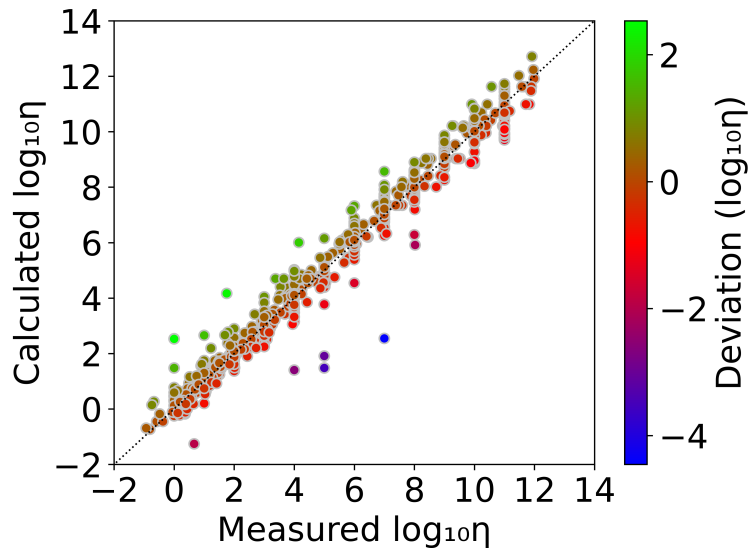

Figure 5. Correlation between predicted values of model of present work and test data.

The skewness and the kurtosis was evaluated by the Fisher method [10], being a distribution with positive or negative skewness and a leptokurtic distribution, respectively. The model of the present work has an skewness of -1.276 and kurtosis of 12.941. The correlation between the values predicted by the model of the present work with the test data (Figure 5) demonstrate a consonance with the line y=x, a line where there is a perfect correlation.

## 4  Conclusion

• Viscosity prediction models ($\eta$ in Pa.s) of lead-containing glasses were built through the artificial neural network algorithm using training and validation data taken from the SciGlass database, with the input variables being the chemical composition composed of the chemical species $SiO_2$, $Al_2O_3$, $B_2O_3$, CaO, $K_2O$, $Na_2O$, PbO (mol%) and temperature (T in K);

- The best artificial neural network contains 26 nodes in the hidden layer (ANN-26), with lowest mean absolute error (MAE) and highest coefficient of determination ($R^2$) in relation to 41 different artificial neural networks;
- Through test data, the results of the MAE and the standard deviation (STD) of the deviation of the models S2, Watt-Fereday, Bomkamp, Lakatos, Riboud, Duchesne, ANNliq and of the best artificial neural network were compared, being the model of the present work with the minimum MAE (0.352 $\log_{10} \eta$) and STD (0.415 $\log_{10} \eta$);
- The sensitivity analysis showed that the model built is in accordance with the literature, indicating that the chemical species $B_2O_3$, $CaO$, $K_2O$, $Na_2O$, $PbO$ decrease viscosity with $SiO_2$, $Al_2O_3$ increasing viscosity, also according to temperature, decreasing viscosity;
- Skewness and kurtosis of the deviations from the values predicted by the model of the present work to the test data were calculated (-1.276 and 12.941, respectively) and the correlation between the values predicted of viscosity by the model of the present work showed agreement with the test data.